\documentclass[sigconf, 9pt, nonacm]{acmart}
\AtBeginDocument{%
  }

\usepackage{algorithm}
\usepackage{algorithmic}
\usepackage{bbold}
\usepackage{subfigure}
\setcopyright{acmlicensed}
\copyrightyear{2025}
\acmYear{2025}
\acmDOI{XXXXXXX.XXXXXXX}
\acmConference[Conference acronym 'EN-IoT]{2nd Workshop on Energy Neutral and Sustainable IoT Devices and Infrastructure}{Sep. 22,
  2025}{KUL, Leuven}
\acmISBN{978-1-4503-XXXX-X/2018/06}

\begin{document}

\title{Evaluating Task Execution Performance Under Energy Measurement Overhead}

\author{Mateen Ashraf}
\email{mateen.ashraf@oulu.fi}
\orcid{1234-5678-9012}
\affiliation{%
  \institution{Center for Wireless Communications, University of Oulu}
  \city{Oulu}
  \country{Finland}
}

\author{Shahab Jahanbazi}
\email{shahab.jahanbazi@oulu.fi}
\affiliation{%
  \institution{Center for Wireless Communications, University of Oulu}
  \city{Oulu}
  \country{Finland}}

\author{Onel L. A. López}
\email{onel.alcarazlopez@oulu.fi}
\affiliation{%
  \institution{Center for Wireless Communications, University of Oulu}
  \city{Oulu}
  \country{Finland}
}

\renewcommand{\shortauthors}{Ashraf et al.}

\begin{abstract}
Energy-awareness for adapting task execution behavior can bring several benefits in terms of performance improvement in energy harvesting (EH) Internet of Things (IoT) devices. However, the energy measurement cost of acquiring energy information, which is traditionally ignored, can potentially neutralize or even reverse the potential benefits. This paper highlights operational parameters, such as energy measurement frequency and task execution frequency, which can be tuned to improve the task execution performance of an EH-IoT device. To this end, we consider energy-blind (EB) and energy-aware (EA) task decision approaches and compare their task completion rate performance. We show that, for specific hardware design parameters of an EH-IoT device, there exists an optimal energy measurement/task execution frequency that can maximize the task completion rate in both approaches. Moreover, if these parameters are not chosen appropriately, then energy measurement costs can cause EA scheduling to underperform compared to EB scheduling.
\end{abstract}

\ccsdesc[500]{Computing methodologies~Simulation evaluation}
\ccsdesc[100]{Hardware~Reusable energy storage}

\keywords{Energy awareness, energy harvesting, energy measurement, Internet of Things, task execution.}


\received{XX June XXXX}
\received[revised]{XX June XXXX}
\received[accepted]{XX June XXXX}

\maketitle
\section{Introduction}
\begin{figure*}[t]
    \centering
    \subfigure[EB scheduling.]
    {%
        \includegraphics[width=0.49\textwidth]{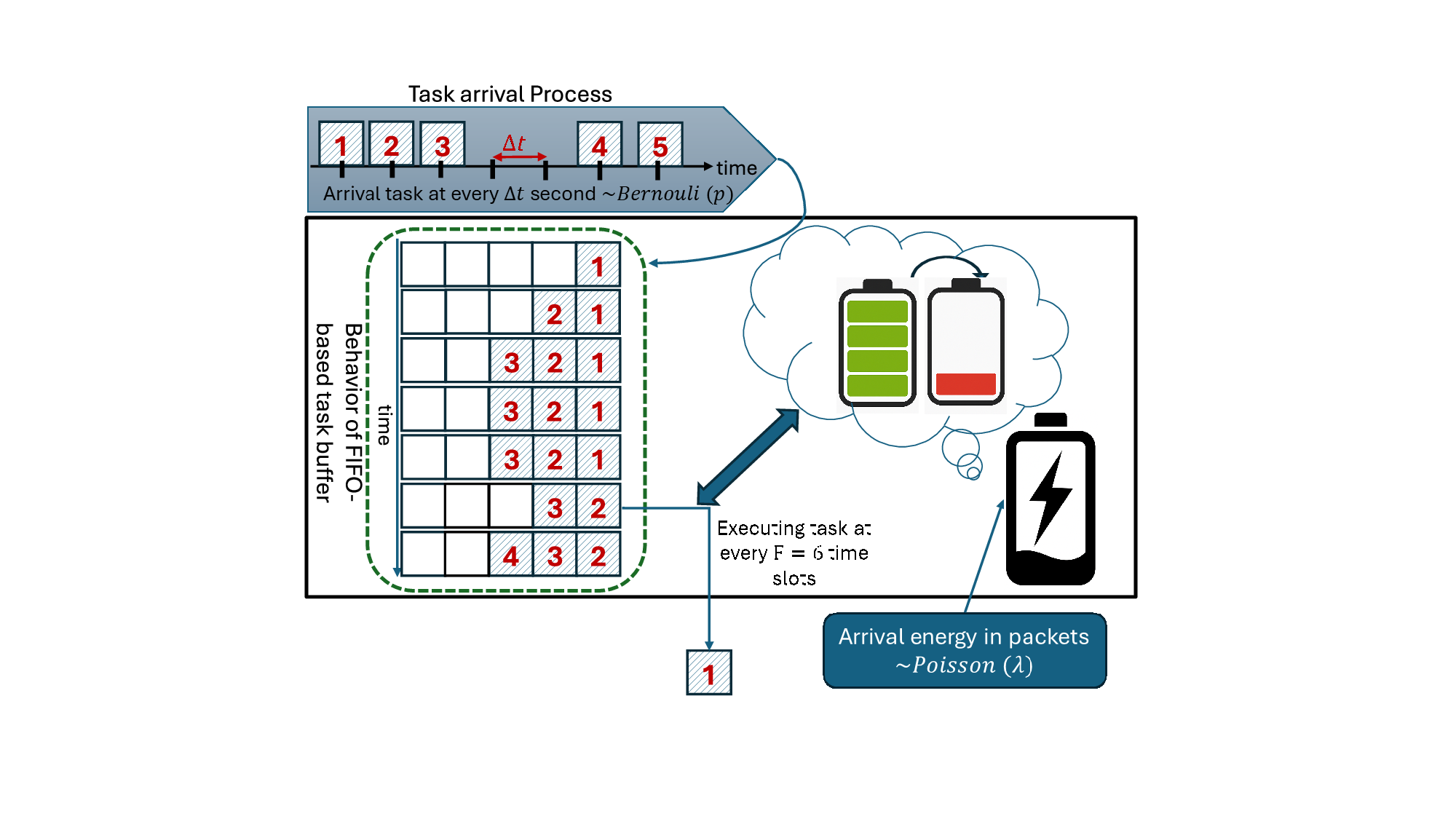}
        \label{fig:system1}
    }
    \subfigure[EA scheduling.]
   {%
        \includegraphics[width=0.49\textwidth]{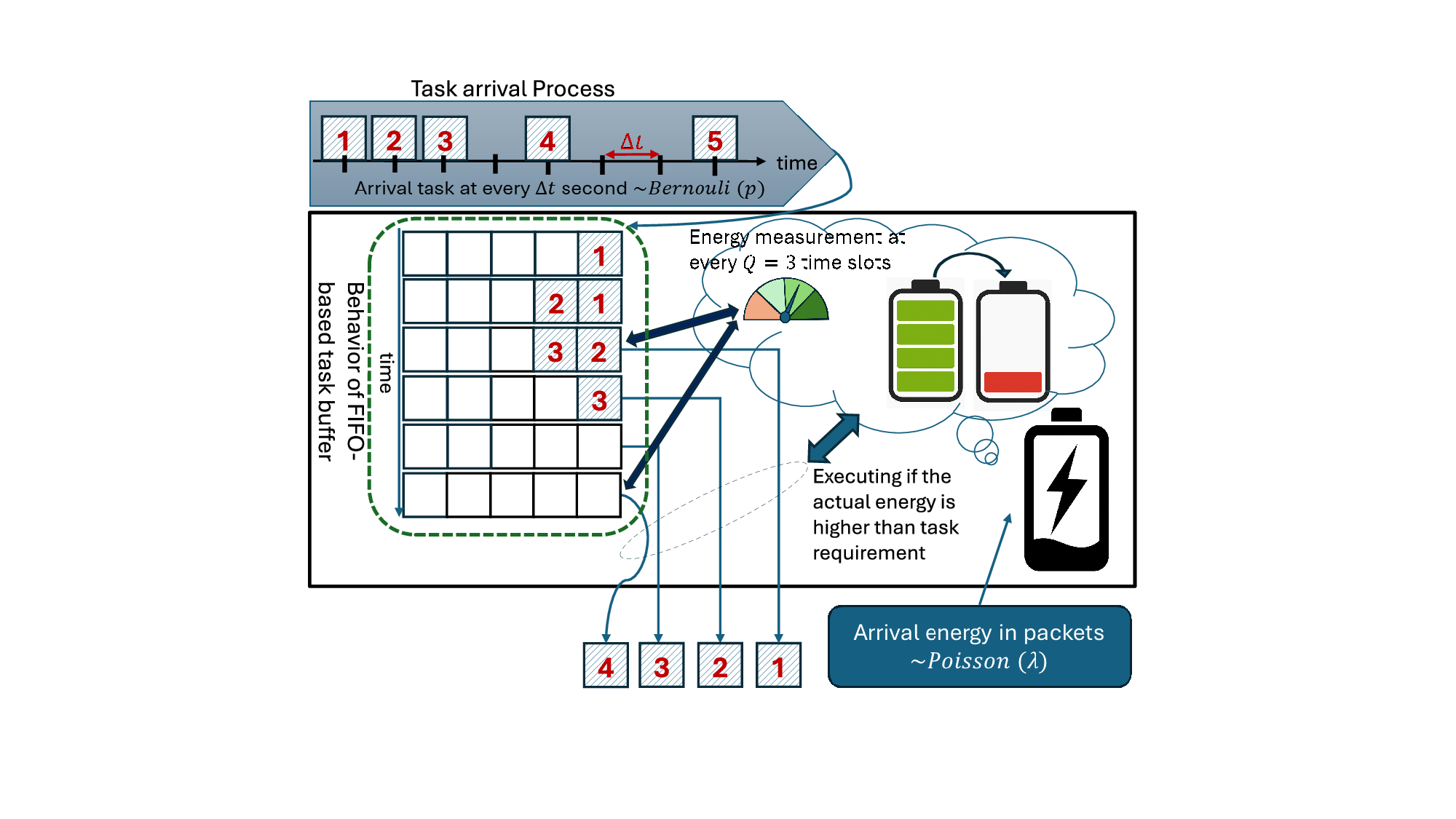}
        \label{fig:system2}
    }
    \caption{EH-IoT device with EB and EA scheduling. (a) EB scheduling: Task execution is attempted every $F$ time slots, with $F=6$ used as example, regardless of the amount of available energy in the energy storage. (b) EA scheduling: Task execution decisions are guided by the periodic energy measurements every $Q$ time slots, with $Q=3$ as an example.}
    \label{fig:system_model}
\end{figure*}
In energy harvesting (EH) Internet of Things (IoT) devices, efficient energy management is critical for reliable operation under highly random and often scarce energy conditions \cite{DMa,UMESH, onel2}. One widely considered approach is to measure or estimate the available energy prior to executing tasks, allowing the device to adapt its task execution decisions to the available energy budget, and thereby reducing the probability of task failure. These schemes, which base task execution decisions on available energy, EH statistics, energy consumption patterns and/or their forecasts, are called energy-aware (EA) schemes. However, it is noted that the measurement process in EA schemes incurs a non-negligible energy overhead \cite{GLius, Ylee}, which can even negate its intended benefits, especially in ultra-low-power devices such as EH-IoT devices.

Energy measurement in EH-IoT devices can generally be classified into two main approaches \cite{9406578}: hardware-based methods and software-based (or model-based) estimation. The former methods involve the direct monitoring of energy levels through sensors and specialized circuitry \cite{BANOTRA}, such as energy monitoring integrated circuits designed especially for low-power applications. These methods typically offer high accuracy but come at the cost of increased system complexity and additional energy consumption. In contrast, the latter approaches rely on mathematical models and observed system behavior to estimate energy usage. Techniques like energy profiling \cite{SABOVIC}, which leverage pre-measured energy consumption patterns of tasks to estimate runtime usage, have less overhead but generally provide lower accuracy. Moreover, the energy profiling methods require higher computational capabilities at the device side to extract information needed for building energy profiles. This can be a serious hurdle in implementation of these methods in EH-IoT devices.

In addition to energy consumption estimation, profiling energy arrival is a crucial factor in enabling an energy-efficient task decision approach for EH-IoT devices. A significant challenge in this context is that the EH profile is typically unknown and highly variable due to environmental fluctuations. To address this, two primary approaches have been developed: prediction-based models \cite{prediction, onel} and stochastic or modeling-based approaches \cite{stochastic_approach}. In prediction-based methods, historical EH data are used to train models that can forecast future energy availability. These techniques often employ time series analysis or machine learning algorithms to anticipate short- or long-term trends. In contrast, modeling-based approaches treat the energy arrival process as a stochastic phenomenon, using probabilistic models (such as Markov processes) to describe and reason about the uncertainty in energy supply. These models are particularly useful for theoretical analysis and for designing robust EA algorithms under uncertainty.

Across both approaches, several common challenges persist, including ensuring measurement accuracy, managing the energy overhead introduced by the measurement process itself, and addressing the latency. Despite its potential benefits in modern EH-IoT devices, the effects of energy measurement on the overall device performance remains underexplored. To consider the trade-offs between energy-awareness and system efficiency, various studies proposed adaptive measurement techniques. For instance, the authors in \cite{dutycycle} used an approach named duty-cycled sensing to reduce measurement overhead while maintaining acceptable accuracy. Therefore, understanding and quantifying the trade-offs introduced by energy measurement is essential for designing robust, self-sustaining IoT systems that maximize task completion rate while minimizing unnecessary energy expenditure. We take a step in this direction in the current work.

This paper highlights the critical system parameters which influence the task completion performance of the EH-IoT device.  Specifically, we consider EB and EA approaches for making decisions about task execution. Herein, the energy-blind (EB) approach adopts a periodic task execution without any energy measurement mechanism, and therefore it does not entail any energy measurement overheads. On the other hand, the EA approach uses available energy estimates for making decision regarding task execution. The energy estimates are based on periodically carried out energy measurements, entailing an energy measurement overhead in the form of measurement energy consumption, it is shown that for a given set of unalterable hardware parameters such as energy storage capacity, task buffer size, and energy measurement circuitry, there exists an optimal task execution frequency for the EB approach, and an optimal value for the energy measurement frequency in the EA approach.

The rest of the paper is organized as follows. The considered system model is discussed in Section 2. Section 3 discusses EB and EA scheduling. The simulation results are presented in Section 4. Finally, the conclusions are provided in Section 5.

\section{System Model}
We consider an EH-IoT device that needs to execute a certain task repeatedly for a long period of time. This can represent an IoT sensor that needs to repeatedly pre-process the reading of an environmental parameter over an indefinite period of time.\footnote{It can be any task such as sensing, computing communicating, and actuating (or a combination thereof).} We assume that the total time is divided into equal time slots, where the duration $\Delta t$ of the individual time slots is at least equal to the time needed for executing the task successfully conditioned on the availability of sufficient energy. This is actually dictated by the operational speed of the hardware equipment, e.g., central processing unit speed when executing computational tasks. The current work does not take task deadline aspects into consideration, while addressing this aspect could serve as a promising direction for future research.

A pictorial representation of the system model is shown in Fig.~\ref{fig:system_model}. The total number of time slots is assumed to be $T_{max}$. For ease of readability, specific details related to the modeling of tasks, energy, and the performance metric are provided in the following subsections.
\subsection{Task Arrival and Storage Model}
It is assumed that at any given time slot, the task arrival is random and follows a Bernouli distribution with parameter $p$. The EH-IoT device is equipped with a task buffer of size $B$ storing non-executed tasks, and it operates in a first-in-first-out (FIFO) manner. The task buffer accepts an arriving task only if it is not already full. On the other hand, if it is full, then the arriving task is dropped and hence it can not be executed in future. In the event that a task is executed successfully, then the executed task is removed from the task buffer and therefore the vacant place becomes available for subsequently arriving task.
\subsection{Energy Models}
The EH-IoT device can store the harvested energy for future use. The energy storage capacity is denoted by $E_{max}$. For modeling EH, we assume that energy arrives in packets, where the number of packets arriving within a time interval follows the Poisson distribution with parameter $\lambda$. The harvested energy in $i$-th time slot is denoted by $E_h^i$.
\subparagraph{Energy consumption occurs as a result of two different operations. \textit{First}, execution of a task within a time slot consumes $E_t$ energy units. In practice, $E_t$ is not precisely known; only an estimate or bound is usually available. Moreover, $E_t$ may vary over time due to factors such as channel fluctuations and temperature gradients. However, for ease of following analysis, we assume that $E_t$ is perfectly known and hence there is no need to employ separate hardware circuity for its estimation/prediction. \textit{Second}, performing energy measurement also consumes energy, in this case $E_m$ units. In general, the energy measurement process has associated time and energy costs. However, here we assume that the time cost is much smaller than $\Delta t$ and hence it can be ignored. Note that the harvested energy due to accumulation of arriving energy packets during each time slot is very low and hence the device may need to wait for several time slots before resuming task execution.} 
\subsection{Performance Metric}
We consider the task completion rate as the performance metric. A task is considered completed if it is executed on its arrival \textit{or} if it is not dropped by the task buffer and hence executed at a later time slot. Then, the task completion rate is defined as the ratio of the number of completed tasks and the number of arrived tasks for any fixed number of time slots. In other words, task completion rate is an indicative of the probability of successful execution of the arriving task.\footnote{The task completion rate is suitable when tasks have similar energy requirements, as it is independent of the arrival sequence. For tasks with varying energy demands, fairness and average execution delay should also be taken into account. Note that the probability of successful execution estimation accuracy increases with the slot count.}

The performance of the considered system depends on various factors, such as the frequency of energy measurements and the energy cost of individual measurements. They can either improve or degrade the task completion rate performance depending on the values of the remaining system parameters. To illustrate these effects, two decision approaches for the execution of tasks are discussed in the following section.
\section{Task Execution Scheduling}
There are two possible approaches to task execution decisions. Either the task execution decisions are EA or EB. Task execution decisions in EA scheduling are based on the energy related expects of the system. These aspects can include, among others (such as EH statistics, energy consumption patterns and/or their forecasts), the state of available energy in the energy storage. On the other hand, the task execution decisions in EB approaches are impervious to the state of available energy. The following subsections explain the operational mechanism of each of these approaches.
\subsection{Energy-Blind (EB) Scheduling}
In this approach, task execution decisions are taken regardless of the available energy state, and therefore no energy measurement cost is incurred in this scheduling. Specifically, in contrast to the EA scheduling, here the EH-IoT decides to execute a task every $F$ time slots without measuring the available energy. Moreover, since task execution times are predetermined and do not depend on available energy, there is no need to calculate an estimate for the available energy.

In this scheduling, a task completion can occur at only every $F$-th time slot if the actual available energy at the beginning of that time slot is sufficient to successfully execute the task. The completed task is then removed from the task buffer. However, if the available energy is not sufficient, then the considered task is not removed from the task buffer and the status of the available energy is changed to zero to reflect energy wastage in the case of unsuccessful execution.

A detailed description of the simulation setup and the working principle of EB scheduling are presented in \textbf{Algorithm 1}. Here, lines 2-8  generate the task and then store it in the task buffer or drop it depending on the current status of the task buffer. The generation of the energy packets and their storage are reflected in lines 9 and 10, respectively. Regarding EB scheduling specific steps, lines 11 and 16 check whether a task can be executed in the current time slot or not, and whether sufficient energy is present or not in the energy storage, respectively. Depending on whether the condition in line 11 or line 16 holds, lines 12-14 or line 17 are used to update the task buffer and the energy state of the storage, respectively.

\begin{algorithm}[t]
\label{algo1}
\caption{Energy-blind scheduling}
\begin{algorithmic}[1]
\STATE Initialize task buffer, energy storage
\FOR{each time slot $t = 1$ to $T_{max}$}
    \STATE Generate task arrival with probability $p$
    \IF{task buffer not full}
        \STATE If task arrived, then add task to FIFO buffer (capped by task buffer size $B$)
    \ELSE
        \STATE If task arrived, then increment dropped task counter
    \ENDIF
    \STATE Generate energy arrival using Poisson process with parameter $\lambda$
    \STATE Update energy storage (capped by capacity $E_{M}$)
    \IF{$t \bmod F = 0$ \AND task buffer not empty \AND actual energy $\geq$ task requirement}
        \STATE Remove task from task buffer
        \STATE Subtract fixed energy for task execution
        \STATE Increment executed task count
    \ELSE
        \IF{$t \bmod F = 0$ \AND task buffer not empty \AND actual energy $<$ task requirement}
        \STATE Change status of stored energy to zero
        \ENDIF
    \ENDIF
\ENDFOR
\end{algorithmic}
\end{algorithm}

\begin{algorithm}[t]
\label{algo2}
\caption{Energy-aware scheduling}
\begin{algorithmic}[1]
\STATE Initialize task buffer, actual and estimated energy levels
\FOR{each time slot $t = 1$ to $T_{max}$}
    \STATE Generate task arrival with probability $p$
    \IF{task buffer not full}
        \STATE If task arrived, then add task to FIFO buffer (capped by task buffer size $B$)
    \ELSE
        \STATE If task arrived, then increment dropped task counter
    \ENDIF
    \STATE Generate energy arrival using Poisson process with parameter $\lambda$
    \STATE Update actual energy (capped by capacity $E_{M}$)
    \IF{$t \bmod Q = 0$}
        \STATE Subtract the energy measurement cost ($E_m$) from the actual energy and update the actual energy
        \STATE Set estimated energy = actual energy
    \ENDIF
    \IF{estimated energy $\geq$ task requirement \AND task buffer not empty}
        \STATE Subtract used energy from the estimated energy
        \IF{actual energy $\geq$ task requirement}
            \STATE Execute task
            \STATE Subtract energy 
            \STATE Remove task from task buffer
            \STATE Increment executed task count
        \ELSE
            \STATE Change status of stored energy to zero
        \ENDIF
    \ENDIF
\ENDFOR
\end{algorithmic}
\end{algorithm}
\subsection{Energy-Aware (EA) Scheduling}
In the EA decision approach, the EH-IoT device measures energy at regular time periods, denoted by $Q$ time slots. Then, in each time slot the choice of executing or deferring the execution of task is made on the basis of most recent energy measurement when energy measurements are not performed in each time slot, it is assumed that the device uses a conservative estimate of the available energy in such time slots. In particular, the estimated available energy in $n$-th time slot post measurement is obtained by subtracting the energy consumed in task executions during $n$ time slots from the most recent measured value. Mathematically, if the actual energy for the $Q\lfloor\frac{n}{Q}\rfloor$-th time slot is denoted by $E_{mea}^{\left(Q\lfloor\frac{n}{Q}\rfloor\right)}$, then the estimated energy for $n$-th time after the latest measurement is given as
\begin{equation}
    E_{est}^{\left(n\right)} = \max\left(E_{mea}^{\left(Q\lfloor\frac{n}{Q}\rfloor\right)} -\sum_{i=1}^{\hat{n}-1}\mathbb{1}^{(i)}E_t,0\right),
\end{equation}
where $\hat{n}=n-Q\lfloor\frac{n}{Q}\rfloor$ and the indicator function $\mathbb{1}^{(i)}$ is $1$ if the task execution takes place in the $i$-th time slot, otherwise it is $0$. On the other hand, the actual energy, denoted by $E_{act}^{(n)}$, in $n$-th time slot is given as
\begin{equation}
    E_{act}^{(n)} = \max\left(E_{mea}^{\left(Q\lfloor\frac{n}{Q}\rfloor\right)} -\sum_{i=1}^{\hat{n}-1}\mathbb{1}^{(i)}E_t+\sum_{i=1}^{\hat{n}-1}E_h^{(i)},0\right).
\end{equation}
In practice, $E_{est}^{(n)}$ is a lower-bound on $E_{act}^{(n)}$ as it assumes no EH between measurements.

In this approach, a task is considered completed only if i) the task execution decision is made based on the estimate of the available energy, and ii) the actual available energy is sufficient to carry out the execution successfully. Subsequently, a task is removed from the task buffer only if it is completed; otherwise, it remains in the task buffer even if an execution decision is made. On the other hand, a task failure occurs when a task arrives and the task buffer is already full due to previously delayed tasks.

\textbf{Algorithm 2} captures the simulation setup and working principle for the EA scheduling. Again, lines 2-8 represent the arrival of the tasks, their subsequent storage in the task buffer, and the possibility of dropping a task. The energy arrival process and the actual energy state update are represented by lines 9 and 10, respectively. The EA scheduling specific steps, such as the periodic energy measurement process and its measurement cost-based energy update is represented by lines 11-14. Finally, contingencies related to task execution, energy state update and task removal from buffer are dealt through lines 15-26.

\subsection{Comparative Discussion}The EB scheduling is less complex than the EA scheduling as there is no dedicated energy measurement circuit in the EB scheduling. Moreover, on a computational level, the EB scheduling does not require an estimation step for estimating the available energy. Meanwhile, their relative performance in terms of the success task completion rate depends heavily on specific system parameters, such as such as $E_M, B, p$, and $\lambda$, and scheduling-specific parameters; these are $Q$ and $E_m$ in the EA scheduling and $F$ in the EB scheduling. For example, when $E_m$ is very small there is almost no energy measurement cost and the EA scheduling can perform measurements frequently. Then, the EH-IoT device can chose to execute the task only when it has sufficient energy thus, decreasing the possibility of wasting energy on non-executable tasks. Consequently, it can be expected that for same values of $Q$ and $F$, the EA scheduling will perform better than the EB scheduling. A detailed discussion about the effects of different system parameters on the performance of both schemes is provided in the following section.
\begin{table}[t]
  \caption{Simulation parameters}
  \vspace{-2mm} 
  \begin{tabular}{ccc}
    \toprule
    Parameter & Description & Value \\
    \midrule
    $p$ & Bernouli distribution parameter & $\{.2, .35, .5\}$ \\ 
    $\lambda$ & Poisson distribution parameter & $\{.25,.5,.75\}$ \\
    $E_m$ & per measurement energy cost & $\{.2, .3, .5\}$ units \\
    $E_t$ & energy for task execution &  $2$ units \\
    $E_M$ & maximum storage capacity & $\{5, 10\}$ units \\
    $Q$ & energy measurement period & $[1,30]$ \\
    $F$ & task execution period & $[1,30]$ \\
    $T_{max}$ & total time slots & $10^5$ \\
    $\Delta t$ & time slot duration & $1$ unit \\
    $B$ & task buffer size & $\{2,10\}$ \\
  \bottomrule
\end{tabular}
\end{table}
\section{Simulation Results}
The simulation parameters are provided in Table 1. Several parameter configurations are tested to assess different operation regimes of the considered schemes. In the following, we illustrate the effects of different system parameters on the task completion performance of the EA and EB scheduling. Specifically, we show the performance of both schemes with respect to task arrival rate, energy arrival rate, task buffer size, and scheduling-specific parameters for each scheme.

\begin{figure}[t]
  \centering

  \subfigure{
    \includegraphics[width=0.5\textwidth]{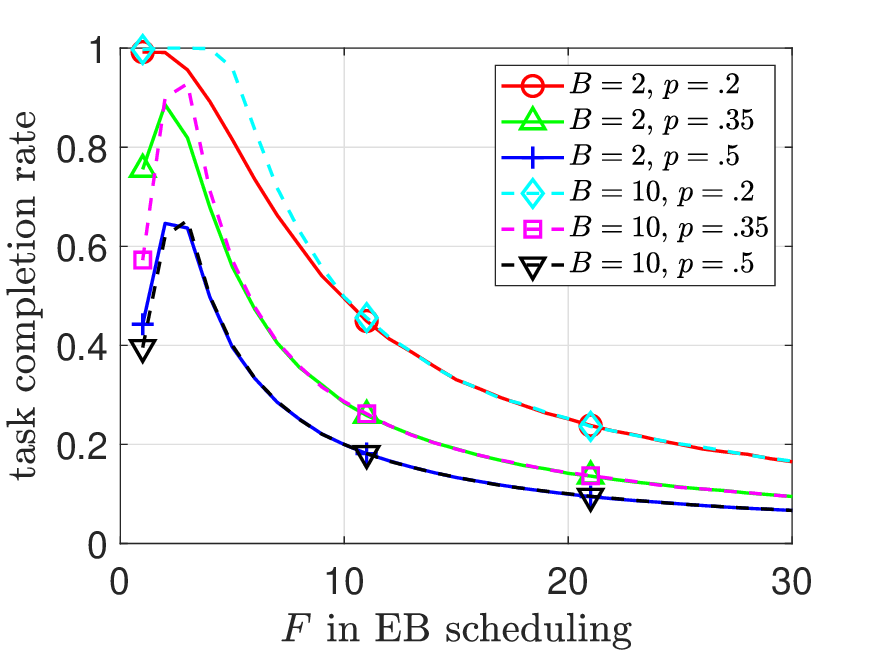}
    \label{Fig1}
  }
  \hfill
  \subfigure{
    \includegraphics[width=0.5\textwidth]{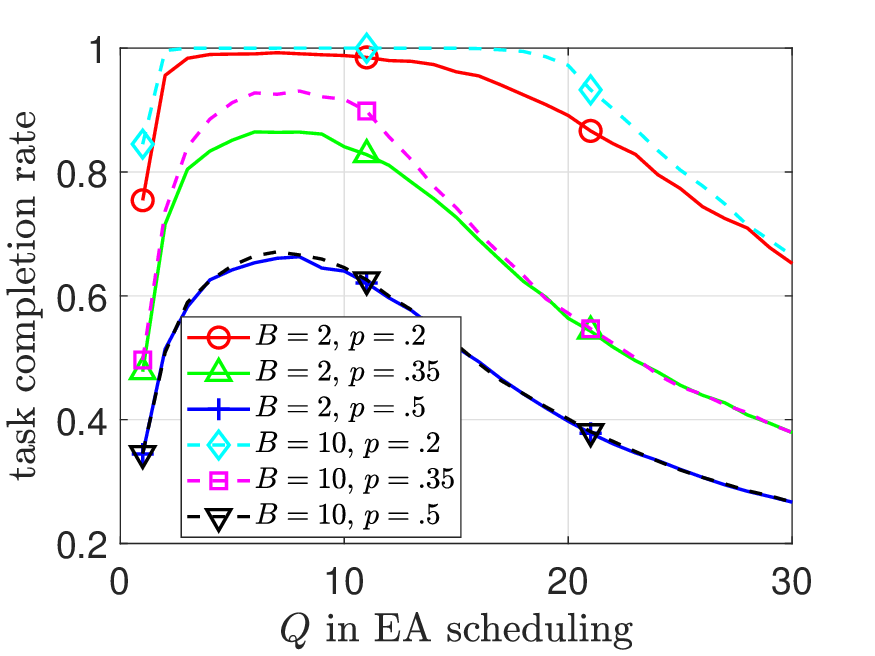}
    \label{Fig2}
  }

  \caption{Task completion rate with respect to task execution period ($F$) in EB (top), and energy measurement period ($Q$) in EA scheduling for different task arrival rates (bottom).}
  \label{fig:scheduling}
\end{figure}

\begin{figure}[t]
  \centering

  \subfigure{
    \includegraphics[width=0.5\textwidth]{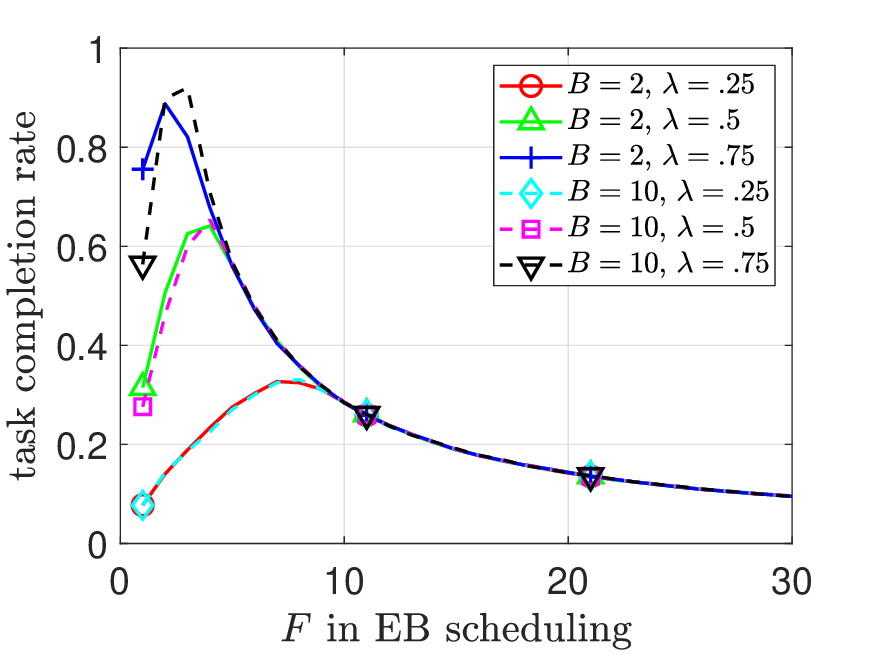}
    \label{Fig3}
  }
  \hfill
  \subfigure{
    \includegraphics[width=0.5\textwidth]{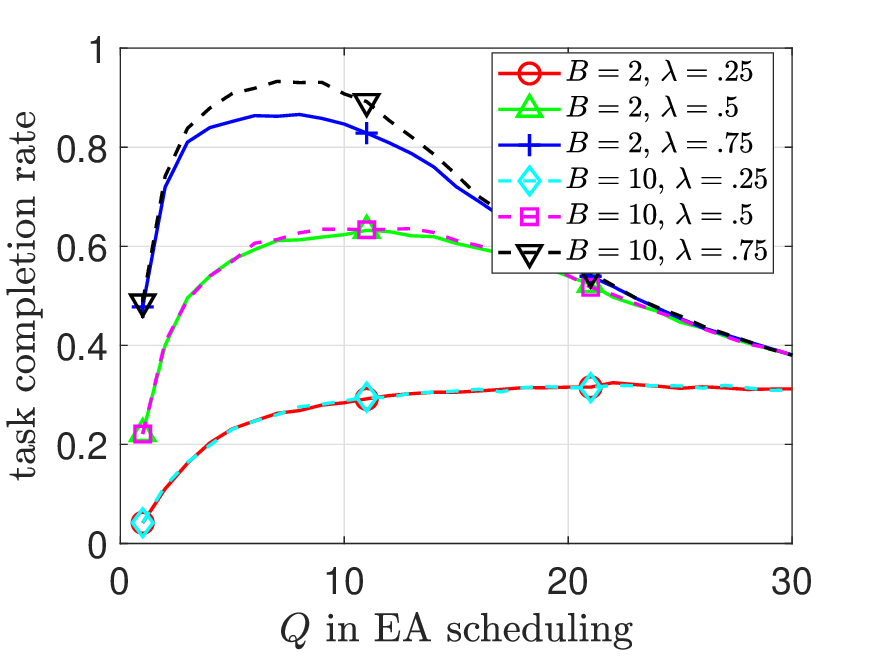}
    \label{Fig4}
  }

  \caption{Task completion rate with respect to task execution period ($F$) in EB (top), and energy measurement period ($Q$) in EA scheduling for different energy arrival rates (bottom).}
  \label{fig:scheduling}
\end{figure}

Fig.~\ref{Fig1} and  Fig. \ref{Fig2} show the task completion rate performance of the EB and EA scheduling with respect to task arrival rate, respectively. It is observed that the performance of both schemes degrades with a task arrival rate increase. This is because a larger $p$ can overburden the system and cause task buffer to remain full for most time slots. This effect can be mitigated by increasing the task buffer size. However, the improvement brought in by the use of increased task buffer size are neutralized if task execution frequency ($F$) in EB scheduling and energy measurement frequency ($Q$) in EA scheduling is increased. In EB scheduling, this is due to the fact that a higher $F$ will cause delays in consecutive task executions, which will result in its increased occupancy rate. This effect will cause any subsequently arriving tasks to be dropped. For EA scheduling, the decrease in performance is caused by the use of conservative energy estimates for all those time slots where energy measurements do not occur. Hence, a lower estimate for the available energy in those time slots causes task executions to be further delayed, thus, resulting in degradation of performance.

Next, Fig. \ref{Fig3} and Fig. \ref{Fig4} show the effects of the energy arrival rate on the performance of the EB and EA scheduling, respectively. Quite expectedly, the performance for both schemes increases with increase in the energy arrival rate. However, we note that for smaller energy arrival rate, the effect of increasing task buffer size is negligible for both schemes. This is because a small energy arrival rate results in more frequent storage depletion, which causes higher occupancy rate of the task buffer, consequently, leading to a higher task drop rate.


\begin{figure}[t]
  \centering

  \subfigure{
    \includegraphics[width=0.5\textwidth]{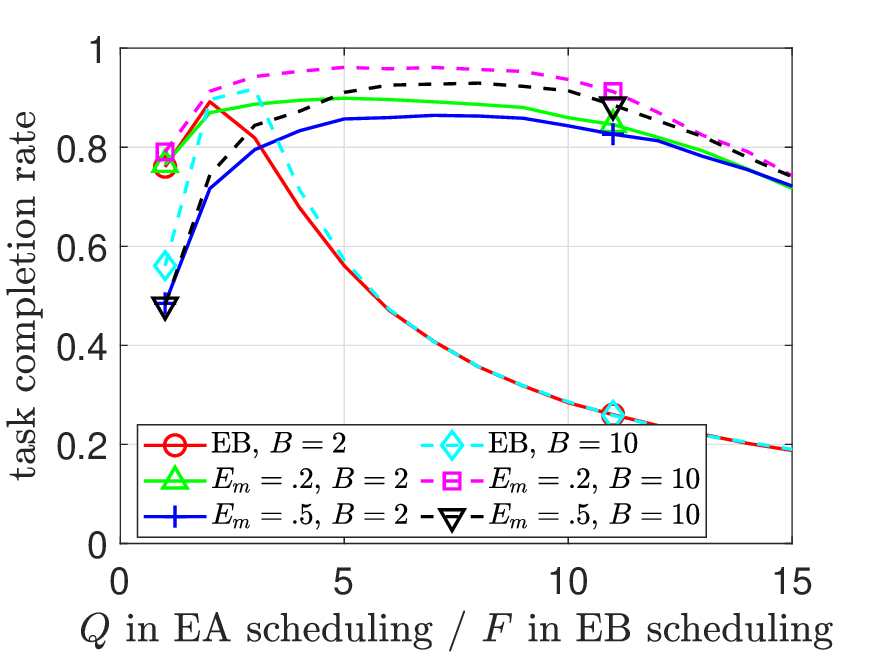}
    \label{Fig5}
  }
  \hfill
  \subfigure{
    \includegraphics[width=0.5\textwidth]{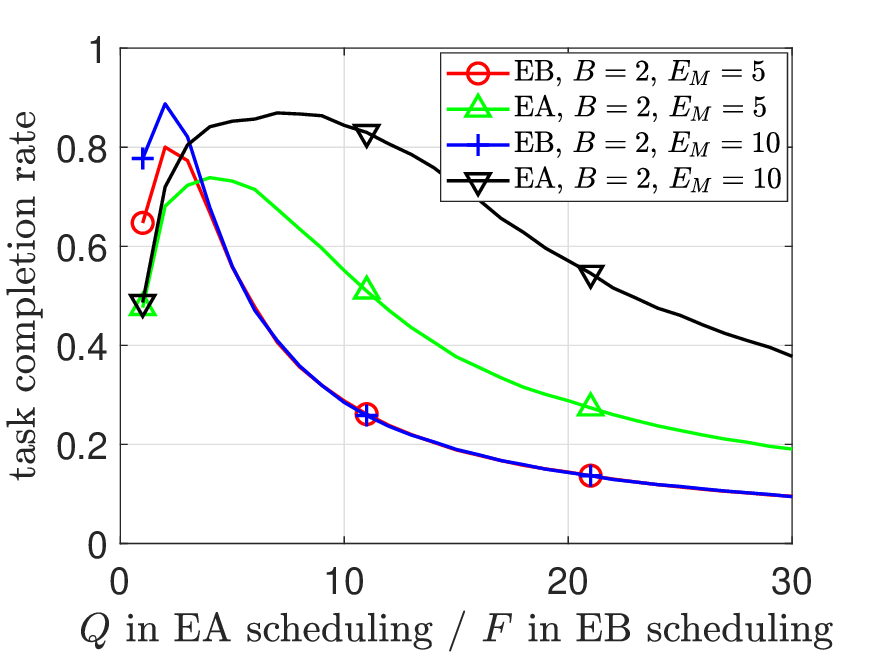}
    \label{Fig6}
  }

  \caption{Task completion rate with respect to task execution period in EB and EA scheduling for different energy measurement cost (top), and energy storage capacity (bottom).}
  \label{fig:scheduling}
\end{figure}

Finally, Fig. \ref{Fig5} and Fig. \ref{Fig6} show a direct performance comparison between EB and EA scheduling. These results illustrate the fact that each of these schemes, under fixed task buffer size ($B$) and energy storage capacity ($E_M$), outperforms the other if its operational parameters, that is $Q$ for EA scheduling and $F$ for EB scheduling, are chosen appropriately. Specifically, we observe that if $F$ and $Q$ have same values, then EB scheduling performs better than EA scheduling. This is due to the fact that a smaller $Q$ means more energy consumption for the measurement process and less energy availability for the execution of the task. On the other hand, if $Q$ is very high, then, due to the use of a conservative energy estimate, task execution is denied, and hence performance degradation occurs. Therefore, there exists a range of $Q/F$ over which each of the schemes performs better than the other. These results emphasize the need to explore optimization techniques to find the optimal values of $F$ and $Q$ in EB and EA scheduling, respectively.


\section{Conclusions and Future Directions}
This paper investigated the task completion rate performance of EB and EA schedulings. It is illustrated that the non-negligible energy cost of the energy measurement process introduces a concave-like behavior in the performance curve with respect to various controllable parameters, such as energy measurement and task execution frequency. Thus, reflecting the need to optimize these parameters to maximize the performance of task execution.

For future work, it would be valuable to explore more realistic multitask execution system models in which the task buffer accounts for task dependencies, and other performance requirements such as fairness among tasks, execution delays, and priority levels. Moreover, exploring non-FIFO based task storage policies can be an interesting future research topic.
\section{Acknowledgments}
This work is partially supported in Finland by the Research Council of Finland (Grants 362782 (ECO-LITE), and 369116 (6G Flagship)); and by the European Commission through the Horizon Europe/JU SNS project AMBIENT-6G (Grant 101192113).

\bibliographystyle{unsrt}
\bibliography{sample-base}

\begin{thebibliography}{10}

\bibitem{DMa}
Dong Ma~et al.
\newblock {Sensing, Computing, and Communications for Energy Harvesting IoTs: A Survey}.
\newblock {\em IEEE Communications Surveys \& Tutorials}, 22(2):1222--1250, 2020.

\bibitem{UMESH}
Sumanth Umesh and Sparsh Mittal.
\newblock A survey of techniques for intermittent computing.
\newblock {\em Journal of Systems Architecture}, 112:101859, 2021.

\bibitem{onel2}
Onel L.~A. López~et al.
\newblock {Zero-Energy Devices for 6G: Technical Enablers at a Glance}.
\newblock {\em IEEE Internet of Things Magazine}, 8(3):14--22, 2025.

\bibitem{GLius}
Luis Gerhorst~et al.
\newblock {EnergyBudgets: Integrating Physical Energy Measurement Devices into Systems Software}.
\newblock In {\em 2020 X Brazilian Symposium on Computing Systems Engineering (SBESC)}, pages 1--8, 2020.

\bibitem{Ylee}
Yoonmyung Lee~et al.
\newblock {A Modular 1 mm$^{3}$ Die-Stacked Sensing Platform With Low Power I$^{2}$C Inter-Die Communication and Multi-Modal Energy Harvesting}.
\newblock {\em IEEE Journal of Solid-State Circuits}, 48(1):229--243, 2013.

\bibitem{9406578}
Chen Guo~et al.
\newblock A survey of energy consumption measurement in embedded systems.
\newblock {\em IEEE Access}, 9:60516--60530, 2021.

\bibitem{BANOTRA}
Atul Banotra~et al.
\newblock {Energy harvesting in self-sustainable IoT devices and applications based on cross-layer architecture design: A survey}.
\newblock {\em Computer Networks}, 236:110011, 2023.

\bibitem{SABOVIC}
Adnan Sabovic~et al.
\newblock {Towards energy-aware tinyML on battery-less IoT devices}.
\newblock {\em Internet of Things}, 22:100736, 2023.

\bibitem{prediction}
Zhenbo Yuan~et al.
\newblock An energy prediction method for energy harvesting wireless sensor with dynamically adjusting weight factor.
\newblock In {\em Algorithms and Architectures for Parallel Processing: 23rd International Conference, ICA3PP 2023, Tianjin, China, October 20–22, 2023, Proceedings, Part VI}, page 465–477, Berlin, Heidelberg, 2023. Springer-Verlag.

\bibitem{onel}
Onel L.~A. López~et al.
\newblock {Energy-Sustainable IoT Connectivity: Vision, Technological Enablers, Challenges, and Future Directions}.
\newblock {\em IEEE Open Journal of the Communications Society}, 4:2609--2666, 2023.

\bibitem{stochastic_approach}
Vijay~S. Rao~et al.
\newblock Optimal task scheduling policy in energy harvesting wireless sensor networks.
\newblock In {\em 2015 IEEE Wireless Communications and Networking Conference (WCNC)}, pages 1030--1035.

\bibitem{dutycycle}
David~E. Ruíz-Guirola~et al.
\newblock Intelligent duty cycling management and wake-up for energy harvesting iot networks with correlated activity.
\newblock In {\em 2024 58th Asilomar Conference on Signals, Systems, and Computers}, pages 1812--1818, 2024.

\end{thebibliography}


\end{document}